\documentclass[sigconf,natbib=true,nonacm, screen]{acmart}

\usepackage{amsmath,amsfonts}
\usepackage{mathtools}
\usepackage{amsthm}
\usepackage{dsfont}
\usepackage{lipsum}
\usepackage{bm,bbm}
\usepackage{bbm}
\usepackage{bbding}

\usepackage{paralist}
\usepackage{multirow}
\usepackage{graphicx}
\usepackage{subfigure}
\usepackage[font=small,labelfont=bf]{caption}
\usepackage{subfloat}
\usepackage{float}
\usepackage{algorithm,algorithmic}
\usepackage{wrapfig}
\usepackage{booktabs}
\usepackage{verbatim}
\usepackage{enumitem}
\usepackage{setspace}
\usepackage{tcolorbox}
\usepackage{nicefrac}
\usepackage{siunitx}
\usepackage{ulem}
\usepackage{natbib}
\usepackage[T1]{fontenc}
\usepackage{microtype}
\usepackage{hyperref}
\hypersetup{
    colorlinks,
    breaklinks,
    linkcolor = blue,
    citecolor = blue,
    urlcolor  = blue,
}

\def \u {\textbf{u}}

\def \E {\mathbb{E}}
\def \x {\mathbf{x}}

\def \D {\mathcal{D}}
\def \z {\mathbf{z}}

\def \O {\mathcal{O}}
\def \R {\mathbb{R}}

\def \Y {\mathcal{Y}}

\def \c {\mathbf{c}}

\def \p {\mathbf{p}}

\def \b {\mathbf{b}}

\def \s {\mathbf{s}}

\def \e {\mathbf{e}}

\def \X {\mathcal{X}}

\def \u {\textsf{u}}

\def \sfi {\textsf{i}}

\usepackage{tikz}

\renewcommand{\hat}{\widehat}

\renewcommand{\emph}{\textit}

\theoremstyle{definition}

\usepackage{mathtools}
\renewcommand{\hat}{\widehat}

\def \E {\mathbb{E}}
\def \D {\mathcal{D}}

\def \R {\mathbb{R}}

\def \X {\mathcal{X}}
\def \Y {\mathcal{Y}}

\def \x {\mathbf{x}}

\def \epsilon {\varepsilon}

\definecolor{DSgray}{cmyk}{0,1,0,0}

\usepackage{prettyref}

\newcommand{\savehyperref}[2]{\texorpdfstring{\hyperref[#1]{#2}}{#2}}
\newrefformat{eq}{\savehyperref{#1}{Eq.~(\ref*{#1})}}
\newrefformat{eqn}{\savehyperref{#1}{Equation~\ref*{#1}}}
\newrefformat{lem}{\savehyperref{#1}{Lemma~\ref*{#1}}}
\newrefformat{lemma}{\savehyperref{#1}{Lemma~\ref*{#1}}}
\newrefformat{def}{\savehyperref{#1}{Definition~\ref*{#1}}}
\newrefformat{line}{\savehyperref{#1}{Line~\ref*{#1}}}
\newrefformat{thm}{\savehyperref{#1}{Theorem~\ref*{#1}}}
\newrefformat{corr}{\savehyperref{#1}{Corollary~\ref*{#1}}}
\newrefformat{cor}{\savehyperref{#1}{Corollary~\ref*{#1}}}
\newrefformat{sec}{\savehyperref{#1}{Section~\ref*{#1}}}
\newrefformat{app}{\savehyperref{#1}{Appendix~\ref*{#1}}}
\newrefformat{appendix}{\savehyperref{#1}{Appendix~\ref*{#1}}}
\newrefformat{assum}{\savehyperref{#1}{Assumption~\ref*{#1}}}
\newrefformat{ex}{\savehyperref{#1}{Example~\ref*{#1}}}
\newrefformat{fig}{\savehyperref{#1}{Figure~\ref*{#1}}}
\newrefformat{alg}{\savehyperref{#1}{Algorithm~\ref*{#1}}}
\newrefformat{rem}{\savehyperref{#1}{Remark~\ref*{#1}}}
\newrefformat{conj}{\savehyperref{#1}{Conjecture~\ref*{#1}}}
\newrefformat{prop}{\savehyperref{#1}{Proposition~\ref*{#1}}}
\newrefformat{proto}{\savehyperref{#1}{Protocol~\ref*{#1}}}
\newrefformat{prob}{\savehyperref{#1}{Problem~\ref*{#1}}}
\newrefformat{claim}{\savehyperref{#1}{Claim~\ref*{#1}}}
\newrefformat{que}{\savehyperref{#1}{Question~\ref*{#1}}}
\newrefformat{op}{\savehyperref{#1}{Open Problem~\ref*{#1}}}
\newrefformat{fn}{\savehyperref{#1}{Footnote~\ref*{#1}}}
\allowdisplaybreaks

\begin{document}

\title[CHIME: A Compressive Framework for Holistic Interest Modeling]{CHIME: A Compressive Framework for Holistic Interest Modeling}

\author{Yong Bai}
\affiliation{%
  \institution{Kuaishou Technology}
  \city{Beijing}
  \country{China}}
\email{baiyonggiggle@gmail.com}
 
\author{Rui Xiang}
\affiliation{%
  \institution{Kuaishou Technology}
  \city{Beijing}
  \country{China}}
\email{reasonxiang@gmail.com}

\author{Kaiyuan Li}
\affiliation{%
  \institution{Kuaishou Technology}
  \city{Beijing}
  \country{China}}
\email{likaiyuan03@kuaishou.com}

\author{Yongxiang Tang}
\affiliation{%
  \institution{Kuaishou Technology}
  \city{Beijing}
  \country{China}}
\email{tangyongxiang@kuaishou.com}

\author{Yanhua Cheng}
\affiliation{%
  \institution{Kuaishou Technology}
  \city{Beijing}
  \country{China}}
\email{chengyanhua@kuaishou.com}

\author{Xialong Liu}
\affiliation{%
  \institution{Kuaishou Technology}
  \city{Beijing}
  \country{China}}
\email{liuxialong2007@sina.com}

\author{Peng Jiang}
\affiliation{%
  \institution{Kuaishou Technology}
  \city{Beijing}
  \country{China}}
\email{jp2006@139.com}

\author{Kun Gai}
\affiliation{%
  \institution{Kuaishou Technology}
  \city{Beijing}
  \country{China}}
\email{gai.kun@qq.com}

\begin{abstract}
Modeling holistic user interests is important for improving recommendation systems but is challenged by high computational cost and difficulty in handling diverse information with full behavior context. Existing search-based methods might lose critical signals during behavior selection. To overcome these limitations, we propose \texttt{CHIME}: A \underline{\textit{C}}ompressive Framework for \underline{\textit{H}}olistic \underline{\textit{I}}nterest \underline{\textit{M}}od\underline{\textit{e}}ling. It uses adapted large language models to encode complete user behaviors with heterogeneous inputs. We introduce multi-granular contrastive learning objectives to capture both persistent and transient interest patterns and apply residual vector quantization to generate compact embeddings. \texttt{CHIME} demonstrates superior ranking performance across diverse datasets, establishing a robust solution for scalable holistic interest modeling in recommendation systems.

% Experimental results demonstrate that \texttt{CLUB} significantly outperforms existing approaches across various scenarios, providing a scalable and effective paradigm for long-term behavior modeling in recommendation systems.

% heterogeneous input, holistic loss
% \texttt{CLUB} leverages pretrained large language models (LLM) and contrastive tasks to effectively capture relationships between long-term behaviors while capturing heterogeneous information. 
% Furthermore, it retains the full behavior sequence by compressing it into a compact embedding.
 
\end{abstract}

%    <concept>
%        <concept_id>10002951.10003317.10003347.10003350</concept_id>
%        <concept_desc>Information systems~Recommender systems</concept_desc>
%        <concept_significance>500</concept_significance>
%        </concept>
%  </ccs2012>

% \ccsdesc[500]{Information systems~Recommender systems}

% \keywords{Recommendation Systems, Ranking Models, User Interest Modeling, Long Sequential User Behavior}

\maketitle

%!TEX root = ../main.tex

\section{Introduction}
User behavior sequences play a crucial role in recommendation systems, since they are the most direct signals of user feedback and contain rich information of user interest~\citep{DBLP:journals/corr/HidasiKBT15, DBLP:conf/aaai/ZhouMFPBZZG19, DBLP:conf/kdd/ZhouZSFZMYJLG18, DBLP:conf/aaai/LyuDHR20, DBLP:conf/kdd/XiaEPBWGJFZZ23, DBLP:journals/corr/abs-2411-09852, CIKM'20/SIM, Sigir'20/UBR4CTR, arxiv'21/ETA, CIKM'22/SDIM, KDD'23/TWIN, CIKM'24/TWINv2, arxiv'24/MARM}. To model user behaviors, there are typically two trends: real-time modeling and long-term modeling. 
Real-time modeling leverages the most recent user behaviors to capture real-time interests, while long-term modeling uses complete historical sequences to identify consistent interests that may not be affected by short-term fluctuations. Although real-time interests of a user can change rapidly, long-term interests often remain relatively stable, offering valuable insights for generating personalized recommendations. For example, when browsing an e-commerce platform, one user may engage with seasonal sales or trending products , but her enduring preferences, such as a consistent interest in a specific product category (e.g., electronics or fashion), serve as more effective indicators for long-term recommendation.

% also mention resource usage
However, long-term behavior modeling in industrial recommendation systems is challenging due to large data scale and strict response latency requirements. Generally, each user can generate massive amounts of interaction data, with sequence lengths often reaching tens of thousands. For example, in e-commerce platforms, active users produce a continuous stream of interactions daily, causing behavior sequences to grow significantly over time. Meanwhile, industrial systems demand real-time responses, which makes it infeasible to process such lengthy sequences entirely during inference even with a shallow one-layer attention model. These challenges underscore the need to develop specialized optimization techniques to enable efficient long-term sequence modeling while meeting the demands of industrial applications.

\begin{figure}[!t]  %子图加并列
	\centering
	\includegraphics[width=\linewidth]{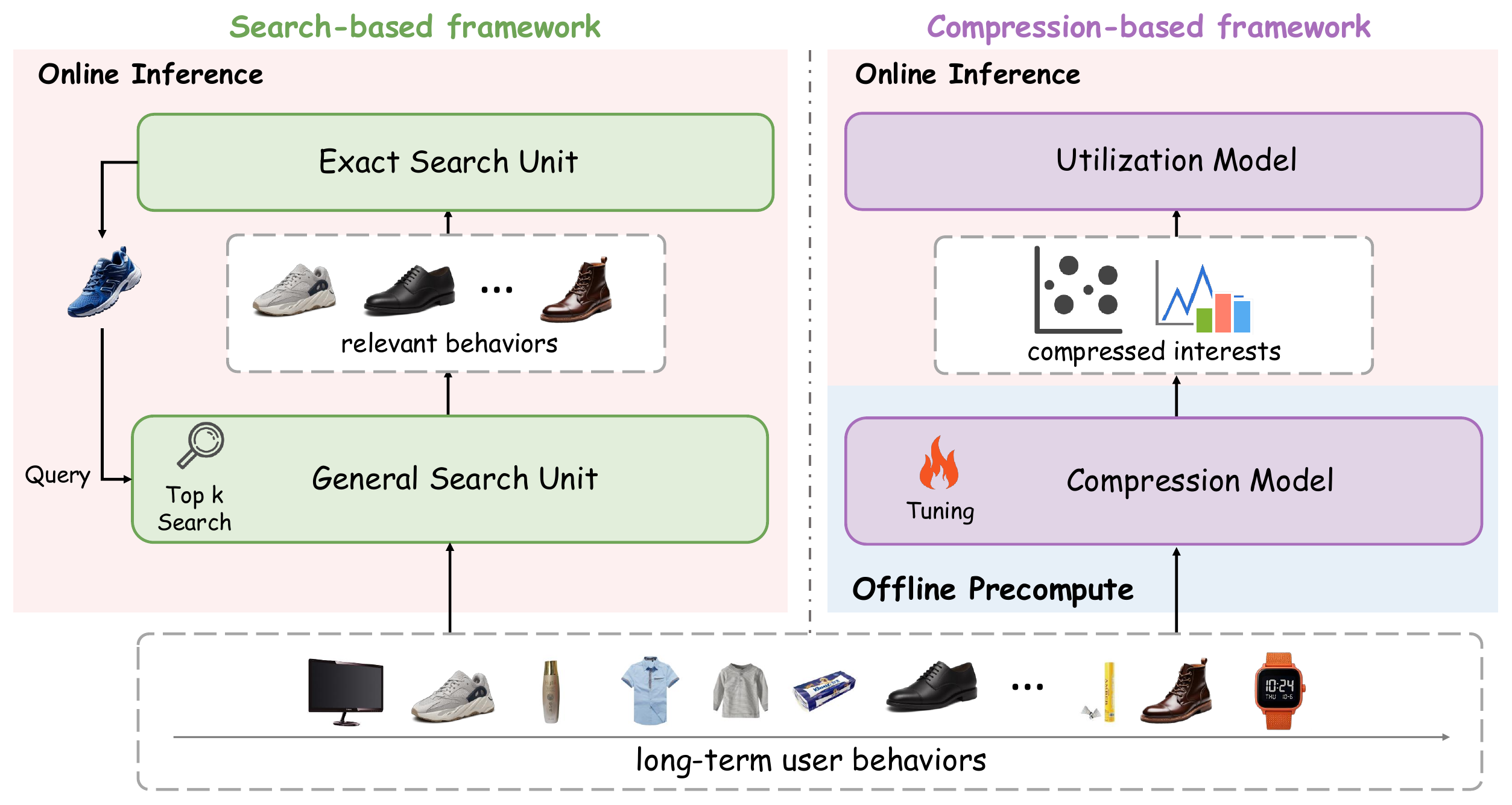}
    \vspace{-5mm}
	\caption{Comparison of search-based and compression-based frameworks: (1)~Effectiveness: The search-based framework relies on simple top-k similarity retrieval, while the compression-based framework leverages richer representations incorporating all behaviors to better capture global interests; (2)~Efficiency: The search-based framework requires per-target online retrieval, whereas the compression-based framework precomputes user representations offline at low frequencies, enabling efficient online serving.}
    \vspace{-5mm}
	\label{fig:comparison}
\end{figure}

Over the past decades, various approaches have been proposed to model long-term behaviors. Among them, one of the noteworthy directions is the
\emph{search-based} two-stage framework~\citep{CIKM'20/SIM, Sigir'20/UBR4CTR, arxiv'21/ETA, CIKM'22/SDIM, KDD'23/TWIN, CIKM'24/TWINv2, arxiv'24/MARM}: (1)\ General Search Unit (GSU) searches a small subset of related behaviors from the long-term sequence based on certain similarity measures, forming a much shorter sequence; (2)\ Exact Search Unit (ESU) employs sophisticated sequence models to model the retrieved shorter sequence with higher precision. 
Despite significantly improving efficiency, search-based methods have a notable limitation: Only the retrieved behaviors actually participate in computation, while a large portion of the behaviors are discarded, potentially leading to suboptimal results.
Other previous studies either employ memory networks to capture behavior interests~\citep{DBLP:conf/sigir/RenQF0ZBZXYZG19, KDD'19/MIMN} or leverage pretrained LLMs to summarize behavior descriptions~\citep{arxiv'24/LIBER, DBLP:conf/recsys/XiLLCZZCT0024}, thereby utilzing the complete long-term behaviors. Nevertheless, these approaches do not implement end-to-end learning across the entire sequence, limiting their capacity to adequately model long-term interests.

In this paper, we aim to design a novel algorithm for long-term behavior modeling that integrates all behaviors into the computational process, thereby capturing holistic long-term interests.
To achieve this, we formulate and investigate a \emph{compression-based} two-stage framework: (1)\ In the first stage, the entire behaviors of one user are fully integrated by a representation module and compressed into a compact low-dimensional representation by a compression module; (2)\ In the second stage, the compressed low-dimensional representation is utilized for further modeling. This is a plug-and-play paradigm that can be easily applied to any ranking model without increasing resource consumption during inference.
% Extensive experimental results show that this method still achieves incremental gains even when applied to the \emph{search-based} algorithms.

Training such long-sequence compression models is challenging due to the time-intensive forward pass and the difficulty of capturing the full range of global user interests, hindering the model's ability to learn from entire behaviors. To address these issues, we leverage the multi-layer causal attention mechanism of LLMs as the \emph{Interest Representation Module (IRM)} and develop a sample construction strategy that enables the computation of thousands of losses in a single forward pass, substantially reducing the computational burden. Furthermore, we introduce two distinct loss functions: \emph{holistic loss}, which captures global interest across the entire sequence, and \emph{immediate loss}, which captures recent interests at specific positions. We also mitigate the risk of overfitting   by utilizing pretrained parameters from LLMs, which help regularize the model and improve its generalization. Our experiments show that models initialized with pretrained LLM parameters consistently outperform those trained from scratch, even when the embedding layer is replaced and only the transformer layers are reused (see Section~\ref{sec-ablation-llm} for experimental results). These findings may provide new insights into the application of pretrained LLMs.

To utilize a pretrained LLM, a natural approach is to convert  behaviors into text input. However, the redundancy of text descriptions means that a single behavior may require multiple tokens, making it difficult to handle long-term behavior sequences. Additionally, the heterogeneity of user behavior information, which may include action data, ID information, and even unstructured text or images, complicates the conversion to text descriptions without omitting crucial information. To address these challenges, we designed the \emph{Interest Adaptation Module (IAM)}, which adapts heterogeneous information to input space of representation module.

To enable downstream ranking models, an effective method is needed to compress representation sequences from the IRM into low-dimensional vectors. A common approach is average pooling, but this method suffers from losing frequency details and failing to capture the complex structures inherent in the data. To address these limitations, we introduce the \emph{Interest Compression Module (ICM)}, which choose to use residual quantization~\citep{DBLP:conf/cvpr/LeeKKCH22} techniques to learn the user interest distribution in an end-to-end manner. The compressed user interest representation is obtained by utilizing a histogram of the learned codebook. Visualization results show that the learned user representation exhibits a satisfactory cluster structure (see Section~\ref{sec-visualization}).

Overall, our main contributions  are summarized as follows:

\begin{enumerate}[leftmargin=5mm]
 \item We propose a novel \emph{compression-based} framework for long-term behavior modeling, enabling to learn a compact representation with the entire behaviors.
\item We design an \emph{Interest Adaptation Module (IAM)} that can end-to-end adapt heterogeneous behavioral information to the input space of LLMs for representation learning.
\item We design an \emph{Interest Representation Module (IRM)} that captures global interests from long sequences and efficiently learns user representations by constructing a holistic loss and an immediate loss. We also investigate the effectiveness of pretrained LLM initialization on this task.
\item We design an \emph{Interest Compression Module (ICM)} that end-to-end learns the user interest distribution and compress it as a compact histogram.
\item Extensive experiments validate the effectiveness of our proposed approach, showing significant improvements across various datasets and tasks, including CTR and CVR tasks in short video, e-commerce, and news recommendation scenarios.
\end{enumerate}

% \vspace{2mm}
% \noindent \textbf{Organization.~~} 
% Section~\ref{sec:related-work} discusses related works. Section~\ref{sec:problem} formulates the problem of long-term behaviors modeling in ranking models. 
% Section~\ref{sec:approach} presents our approach.  Section~\ref{sec:experiment} reports the experimental results. Section~\ref{sec:conclusion} concludes the paper.

%!TEX root = ../main.tex

\section{Related Work}
\label{sec:related-work}
% \vspace{1mm}

\begin{figure*}[!t]  %子图加并列
	\centering
	\includegraphics[width=\linewidth]{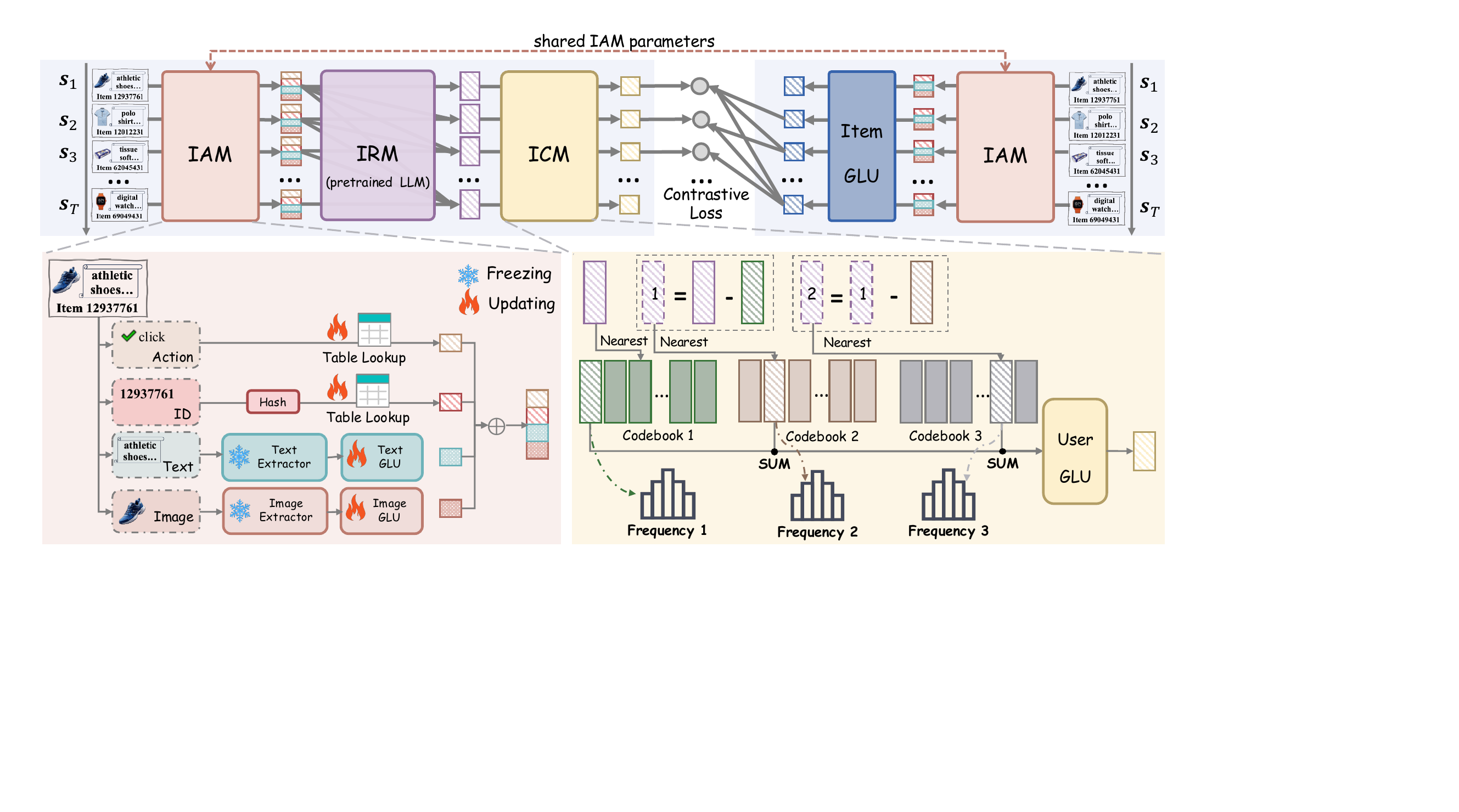}
    \vspace{-3mm}
	\caption{Framework of the compression model, comprising three modules: (1) Interest Adaptation Module (IAM),  integrating heterogeneous information; (2) Interest Representation Module (IRM), tuning a pretrained LLM backbone to capture long-term interests through holistic and immediate contrastive losses; and (3) Interest Compression Module (ICM), compressing interest distribution with residual quantization.}
    \vspace{-3mm}
	\label{fig:framework}
\end{figure*}

\textbf{Ranking Model.} 
Ranking models are a crucial component of recommendation systems, aimed at accurately predicting the relevance of candidate items to users. Common tasks in ranking models mainly include Click-Through Rate (CTR)~\citep{DBLP:conf/cikm/ShengZZDDLYLZDZ21, DBLP:conf/kdd/WuBCRXHDZ22, DBLP:conf/cikm/ZhangCXBHDZ22} and Conversion Rate (CVR)~\citep{DBLP:conf/kdd/ChanZHBSLHLJXZ23, DBLP:conf/aaai/YangLHZZZT21, DBLP:conf/kdd/GuSFZZ21} prediction. Early ranking models predominantly rely on hand-crafted features combined with traditional machine learning algorithms, such as Logistic Regression (LR) and Factorization Machines (FM)~\citep{DBLP:conf/icdm/Rendle10, 10.1145/2959100.2959134}.
With the advent of deep learning, ranking models have significantly evolved, leveraging neural networks to capture complex user-item relationships~\citep{ DBLP:conf/recsys/Cheng0HSCAACCIA16, DBLP:conf/ijcai/GuoTYLH17, DBLP:conf/kdd/LianZZCXS18, DBLP:conf/ijcai/XiaoY0ZWC17, DBLP:conf/www/WangSCJLHC21, DBLP:conf/kdd/ZhouZSFZMYJLG18, DBLP:conf/aaai/LyuDHR20, DBLP:conf/kdd/XiaEPBWGJFZZ23, DBLP:journals/corr/abs-2411-09852}. More recently, the rapid progress of large language models (LLMs) has sparked a wave of studies exploring their application in ranking models~\citep{DBLP:journals/corr/abs-2306-05817, DBLP:journals/corr/abs-2412-13432, DBLP:journals/www/WuZQWGSQZZLXC24, DBLP:journals/corr/abs-2410-19744}.

\vspace{1mm}
\noindent \textbf{Behavior Modeling.} Recent advancements in user interest modeling for ranking tasks have significantly leveraged user behaviors, achieving notable success. Early approaches focus on using deep neural networks to automatically capture user interests from short-term behaviors, employing various architectures such as recurrent neural network (RNN)~\citep{DBLP:journals/corr/HidasiKBT15, DBLP:conf/aaai/ZhouMFPBZZG19}, target-attention~\citep{DBLP:conf/kdd/ZhouZSFZMYJLG18, DBLP:conf/aaai/LyuDHR20} and transformer~\citep{DBLP:conf/kdd/XiaEPBWGJFZZ23, DBLP:journals/corr/abs-2411-09852}. While short-term behaviors have been effectively modeled, long-term behavior modeling is essential for capturing the evolution of user interests over time. Memory-based methods~\citep{DBLP:conf/sigir/RenQF0ZBZXYZG19, KDD'19/MIMN} use memory networks to capture long-term interests. 
Subsequently, SIM~\citep{CIKM'20/SIM} and UBR4CTR~\citep{Sigir'20/UBR4CTR} adopt a two-stage cascaded framework, which first retrieves the most relevant behaviors from long-term behaviors and then  performs complex modeling based on the retrieved behaviors. Building upon the SIM framework, many subsequent works have enhanced the GSU for improved performance~\citep{CIKM'22/SDIM, arxiv'21/ETA, KDD'23/TWIN, CIKM'24/TWINv2, arxiv'24/MARM, arxiv'24/DARE}.

% RNN
% Session-based recommendations with recurrent neural networks
% Deep interest evolution network for click-through rate prediction

% target Attention
% Deep match to rank model for personalized click-through rate prediction
% Deep interest evolution network for click-through rate prediction. 

% Transformer 
% Deep session interest network for click-through rate prediction
% BERT4Rec: Sequential recommendation with bidirectional encoder representations from transformer. 
% TransAct
% InterFormer
%  Behavior sequence transformer for e-commerce recommendation in alibaba.
% ERT4Rec: Sequential recommendation with bidirectional encoder representations from transformer. 
% 

\vspace{1mm}
\noindent \textbf{Pretrained LLMs for Behavior Modeling.} Pretrained large language models (LLMs) have been extensively explored in the recommendation domain~\citep{DBLP:journals/corr/abs-2306-05817, DBLP:journals/corr/abs-2412-13432, DBLP:journals/www/WuZQWGSQZZLXC24, DBLP:journals/corr/abs-2410-19744}. Beyond approaches that avoid fine-tuning LLM parameters by reusing the model architecture for training transformers from scratch~\citep{DBLP:journals/corr/abs-2411-10057, DBLP:conf/icml/ZhaiLLWLCGGGHLS24, arxiv'24/LEARN, DBLP:journals/corr/abs-2502-18965, DBLP:conf/kdd/WangXHZ0LLLXZD24} or leveraging prompt tuning~\citep{arxiv'24/LIBER, DBLP:conf/cikm/SunSZ000024, DBLP:conf/www/LinSZDCQTY024, DBLP:conf/recsys/XiLLCZZCT0024}, most works focus on supervised fine-tuning to align the LLMs with domain-specific knowledge in advertising and recommendation tasks. A widely adopted strategy involves converting recommendation tasks into pure textual descriptions and fine-tuning the LLMs in the text domain~\citep{NAACL'24/Align, DBLP:journals/corr/abs-2409-12740, DBLP:journals/corr/abs-2401-13870, DBLP:conf/www/ZhengCQZ024}. However, given the large semantic space of items and the redundancy in textual descriptions, some works introduces out-of-vocabulary (OOV) tokens to inject both semantic and collaborative information into LLMs, thereby enhancing their capability to model user behavior effectively~\citep{DBLP:conf/icde/ZhengHLCZCW24, DBLP:journals/corr/abs-2310-19488, DBLP:conf/sigir/TanXHGLZ24, DBLP:journals/corr/abs-2406-08477, DBLP:conf/www/ZhangWZXGHC24, DBLP:conf/acl/ZhangBYWF024}.
%!TEX root = ../main.tex

\section{Problem Formulation}
\label{sec:problem} 
The objective of ranking models is to solve a classification problem with a binary label space $\Y = \{-1, 1\}$ and a heterogeneous feature space $\X$. During training, the learner uses historical logs to collect a number of samples from the data distribution $\D(\x,y)$ and learn a model $f:\X\mapsto\Y$ by minimizing the risk function $\E_{\D}\left[\ell(f(\x),y)\right]$, with $\ell(\cdot)$ denoting the classification loss. Once trained, the model $f$ is deployed for real-time inference.

The practical meaning of the label space $\Y$ varies by task. For example, it indicates if a user clicks for CTR prediction and if a user converts for CVR prediction. The feature space $\X$ includes three types: (1) categorical features; (2) numerical features; and (3) sequential features, which represent historical behaviors. A sequential feature of user $\u$, denoted as $\x_{\text{s}}^\u=\{\s_t^\u\}_{t=1}^{T_\u}$, consists of behavior attributes $\s$, where $T_\u$ is the sequence length. In long-term behavior modeling tasks, the sequence length $T$ can be quite large~\citep{CIKM'24/TWINv2} (e.g., $T > 10^3$, potentially reaching $10^6$). For simplicity, we omit the user index notation $\u$ in the following. 

%!TEX root = ../main.tex

\section{Proposed Approach}
\label{sec:approach}
This section introduces the proposed \texttt{CHIME} approach. 
Section~\ref{sec:framework} presents the overall framework. Section~\ref{sec:adaptation},~\ref{sec:representation} and~\ref{sec:compression} details the proposed modules. Section~\ref{sec:efficiency} discusses the efficiency.

\subsection{Overall Framework}
\label{sec:framework}
As shown in Fig.~\ref{fig:comparison}, \texttt{CHIME} comprises two stages: \emph{compression} and \emph{utilization}. During \emph{compression}, the model learns from entire long-term behaviors to generate a low-dimensional representation. During \emph{utilization}, the representation is then used to train ranking models with additional features.

\vspace{1mm}
\noindent\textbf{Compression Stage.} The compression stage trains a model to map the behavior sequence $\{\s_t\}_{t=1}^{T}$ to a low-dimensional representation. 
As shown in Figure~\ref{fig:framework}, the sequence is first processed by the \emph{Interest Adaptation Module} (IAM) to effectively utilize heterogeneous behavioral information and generate user and item inputs.
The user inputs are then fed into the \emph{Interest Representation Module} (IRM), a transformer initialized with pretrained LLMs, and subsequently passed through the \emph{Interest Compression Module} (ICM) to obtain the user hidden states. 
During training, the item inputs are processed by the item gated linear unit (GLU), yielding item hidden states whose dimensions match those of the user hidden states. 
The \emph{holistic loss}, capturing global interests, and the \emph{immediate loss}, capturing recent interests, are then calculated based on user and item hidden states. All modules are designed to be trained in an end-to-end manner.
During inference, the compressed representation is derived from the histogram of the codebook in the \emph{Interest Compression Module}.

\vspace{1mm}
\noindent\textbf{Utilization Stage.} During  utilization, the compressed representations serve as numerical features for training ranking models. The compression model 
operates independently and requires infrequent updates (e.g., daily or weekly), eliminating the need for real-time processing and thereby enhancing computational efficiency. Furthermore, the compressed representations are plug-and-play, allowing seamless integration with various sequence modeling methods, including search-based long-term behavior modeling methods. Experiments in Section~\ref{sec:performance} demonstrate that incorporating the compressed representations with other search-based methods can yield significant performance boosts.

\subsection{Interest Adaptation}
\label{sec:adaptation}
Behavior attributes $\s_t$ can be heterogeneous, encompassing variable categorical information (e.g., item IDs, category IDs, action labels) and unstructured data (e.g., images, texts). 
Previous research primarily uses two approaches: converting all behaviors into text to reuse the tokenizer of LLMs or applying vector quantization to transform semantical attributes into out-of-vocabulary tokens. Whereas, the former fails to effectively process non-textual data, while the latter may lose detailed information during quantization.

In this paper, we propose an Interest Adaptation Module capable of flexibly integrating various modalities. 
As shown in Figure~\ref{fig:framework}, the module comprises two phases: During transformation, action and ID features are embedded via a lookup table. Text and image features are processed through frozen pretrained models and a GLU-based projector to obtain their embeddings.
During concatenation, all embeddings are concatenated to form the user input, denoted by $\{\e_t^{\u}\}_{t=1}^{T}$. And all embeddings except the action labels are concatenated to form the item input, denoted by $\{\e_{t}^{\sfi}\}_{t=1}^{T}$. Dimensions of each attribute are meticulously designed to  align with the input embedding dimension of the LLM.

\subsection{Interest Representation}
\label{sec:representation}
% \vspace{2mm}
\noindent\textbf{Sample Construction.} To train a behavior compression model, a natural way is to pair each candidate item $\s_{t_k}$ and its historical behavior sequence $\{\s_t\}_{t=1}^{t_k - 1}$ to form a single sample. 
However, the extensive input lengths render each forward pass computationally intensive, leading to inefficiency as behavior sequence of a user necessitates $T_\u$ forward passes per epoch.
An alternative approach~\citep{KDD'22/PinnerFormer, arxiv'24/LEARN} sets a time threshold $t_{\tau}$ and constructs losses between historical behaviors $\{\s_t\}_{t=1}^{t_{\tau}}$ before the threshold and future behaviors $\{\s_t\}_{t=t_{\tau}}^{T}$ after the threshold. Whereas, this approach lacks direct constraints within historical and future behaviors, leading to suboptimal sample utilization.
To overcome this limitation, we leverage the causal attention mechanism of LLMs, which employs masks to ensure each token in a sequence attends only to itself and preceding tokens, thereby preventing future information leakage and facilitating autoregressive modeling. 
Each behavior $\s_{t_k}$ aggregates information from all preceding behaviors $\{\s_t\}_{t=1}^{t_k}$, and contrastive loss is calculated based on all subsequent behaviors $\{\s_t\}_{t=t_k}^{T}$, which allows the entire behavior sequence to undergo a single forward pass, enhancing efficiency.

\vspace{2mm}
\noindent\textbf{LLM as Backbone.} 
For each user, the embedding sequence $\{\e_{t}^{\u}\}_{t=1}^{T}$ is fed into a decoder-only transformer, which typically comprises multiple layers with causal attention and a gated linear unit. 
The natural approach to obtain a compression model with a transformer is to train it from scratch with random initialization, but it is challenging since large number of parameters can lead to slow training and overfitting. To mitigate these issues, we propose reusing pretrained LLM parameters, which have strong convergence properties after training on trillions of tokens.
 
A common approach to reuse pretrained LLM parameters is converting behavior sequences into texts~\citep{NAACL'24/Align, arxiv'24/LIBER}. However, this is inefficient for long sequences as the redundancy of texts requires multiple tokens per behavior. Furthermore, some items, like music or short videos, lack textual information. Instead, we adopt a novel approach: discarding the tokenizer and embedding layer of LLMs while retaining only decoder parameters. Our findings reveal that initializing with pretrained decoder parameters of LLMs, even with a randomly initialized embedding layer, outperforms training from scratch (see Section~\ref{sec-ablation-llm}). We conjecture that pretrained parameters act as a form of regularization during training, potentially offering valuable insights for future research on LLMs.

\vspace{2mm}
\noindent\textbf{Loss Functions.} 
To improve sample efficiency and balance global and local behavior relationships, we design two contrastive losses: \emph{holistic loss} and \emph{immediate loss}.  
The holistic loss captures global relationships between behaviors with a softmax-based InfoNCE loss~\citep{DBLP:conf/cvpr/He0WXG20} by treating all future positive behaviors of the same user as positives, and all future negative behaviors, along with the behaviors of other users, as negatives. 
In contrast, the immediate loss focuses on local relationships by applying a sigmoid-based contrastive loss~\citep{DBLP:conf/iccv/ZhaiM0B23} with the next behavior.

Specifically, after feeding the user input $\{\e_{t}^{\u}\}_{t=1}^T$ into a pre-trained LLM and then pass through a user GLU, we obtain a user hidden sequence $\{\z_t\}_{t=1}^T$, where $\z_t$ is a low-dimension vector. Meanwhile, the item input $\{\e_{t}^{\sfi}\}_{t=1}^T$ passes through an item GLU, yielding the item hidden sequence $\{\p_t\}_{t=1}^T$, where vector $\p_t$ matches the dimension of $\z_t$. Let $\{y_t\}_{t=1}^T$ denote the label sequence, where $y_t \in \{-1, 1\}$ means whether the user make a positive action at position $t$. 
The loss can be computed for each position $t$ of each user $\u$ in the batch:
% \vspace{-4pt}
\begin{equation}
    \mathcal{L}_{\text{contrast}} = 
\underbrace{\frac{1}{N_\text{h}} \sum_{\u \in \mathcal{U}} \sum_{t=1}^{T_{\u}-1} \mathcal{L}_{\text{h}}\left(\z_{t}^{\u}\right)}_{\text{(a) holistic loss}} + 
\underbrace{\frac{1}{N_\text{i}} \sum_{\u \in \mathcal{U}} \sum_{t=1}^{T_{\u}-1} \mathcal{L}_{\text{i}}\left(\z_{t}^{\u}\right)}_{\text{(b) immediate loss}},
\label{equ:loss}
\end{equation}
% \vspace{-1pt}
where $\mathcal{U}$ denotes the set of users in a batch. $N_\text{h}$ and $N_\text{i}$ represents the total number of pairs used to compute the loss, respectively.

\begin{figure}[!t]  %子图加并列
	\centering
	\includegraphics[width=.7\linewidth]{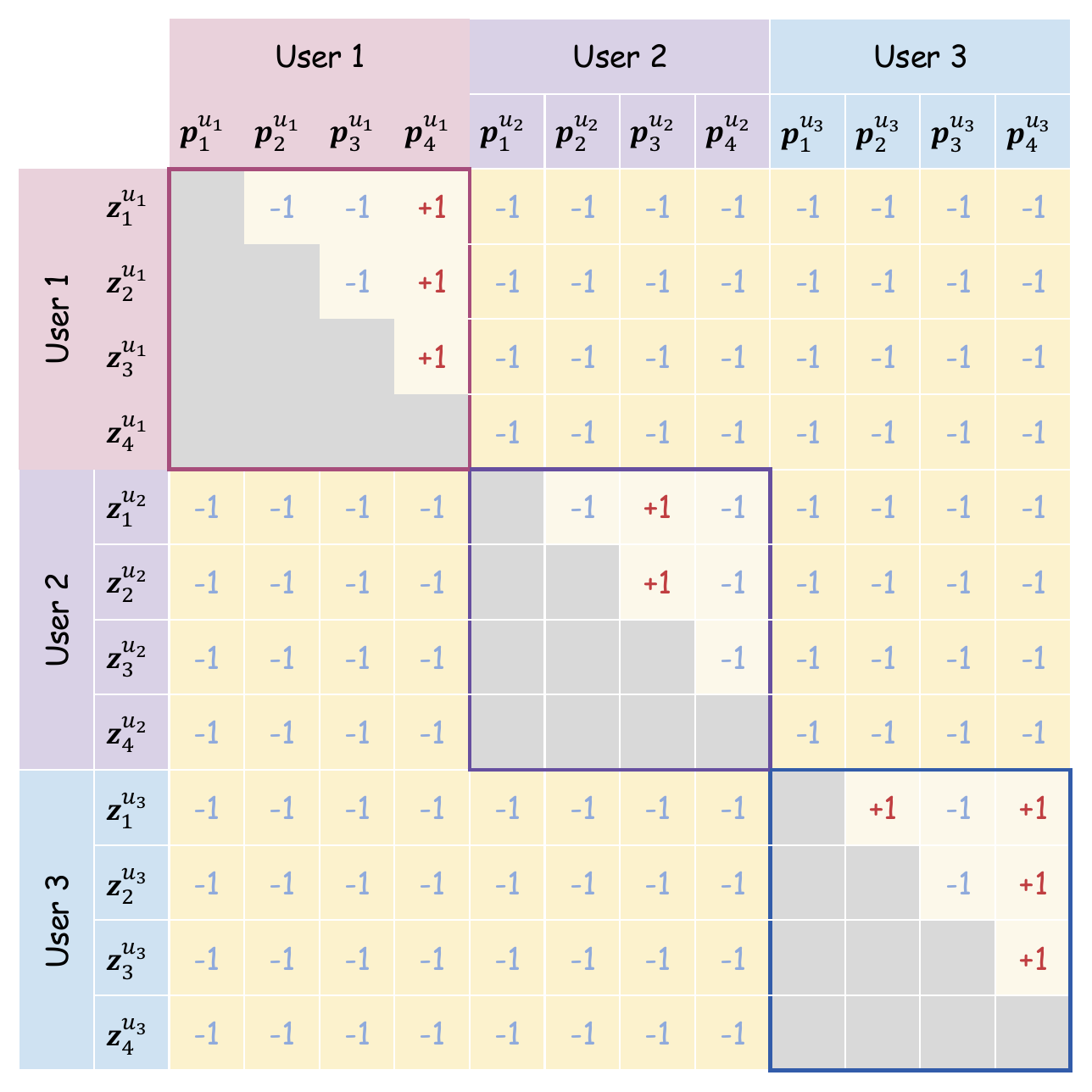}
    \vspace{-3mm}
	\caption{Illustration of holistic loss. Behaviors from other users are considered negative samples. Future behaviors are taken as positive or negative samples based on their labels.}
    \vspace{-5mm}
	\label{fig-holistic}
\end{figure}

The holistic loss (term (a)) in~\eqref{equ:loss} is designed as a softmax-based InfoNCE loss, denoted by
% \vspace{-1pt}
\begin{equation*}
\mathcal{L}_{\text{h}}\big(\z_{t}^{\u}\big) = 
    -\sum_{\p^{+} \in \mathcal{Z}_{\u, t}^{+}} \log \frac{
        \exp\big[s(\z_{t}^{\u}, \p^{+})\big]
    }{
        \sum_{\p \in \mathcal{Z}_{\u, t}^{+} \cup \mathcal{Z}_{\u, t}^{-}} \exp\big[s(\z_{t}^{\u}, \p)\big]
    },
    \label{equ:holistic}
\end{equation*}
% \vspace{-1pt}
where $\mathcal{Z}_{\u, t}^{+}$ and $\mathcal{Z}_{\u, t}^{-}$ represents the item vector sets of the positive samples and negative samples, respectively. The score function $s(\z, \p)$ in~\eqref{equ:holistic} is denoted by $s(\z, \p) = \langle\z, \p\rangle / \tau$, where $\langle \cdot,\cdot \rangle$ is the inner product function and $\tau$ is the given temperature hyper-parameter.
% As shown in Figure X, 
As shown in Figure~\ref{fig-holistic}, for the behavior representation of user $\u$ at position $t$, the positive sample set $\mathcal{Z}_{\u, t}^{+}$ includes all items with $y_t^{\u} = 1$ within the time range $(t, T_{\u})$. The negative sample set $\mathcal{Z}_{\u, t}^{-}$ consists of all items with $y_t^{\u} = -1$ within $(t, T_{\u})$, as well as items from other users within the same batch. The past behavior of the use in the interval $[1, t]$ is not used to construct the loss.

The immediate loss (term(b)) in~\eqref{equ:loss} is defined as a sigmoid constractive loss, denoted by
% \vspace{-1pt}
\begin{equation*}
\mathcal{L}_{\text{i}}\left(\z_{t}^{\u}\right) = -\log \frac{1}{1 + \exp\left[y_{t+1}^{\u}\left(-s\left(\z_{t}^{\u}, \p_{t+1}^{\u}\right)+b\right)\right]},
\end{equation*}
% \vspace{-1pt}
where $\p_{t+1}^{\u}$ and $y_{t+1}^{\u}$ are the item vector and label at time $t+1$, and $b$ is the bias term added to the similarity score for adjustment.

\subsection{Interest Compression}
\label{sec:compression}
% rq layer, mentioning how to represent a user 
To compress long-term interests of users, another key challenge is representing these diverse interests as a low-dimensional vector. Given that user interests often exhibit multiple peaks, for example, on a short video platform, one user may focus on topics such as fitness tutorials and travel vlogs, while another may primarily engage with content like cooking recipes and technology reviews. We model the distribution of user interests with an interest histogram, which captures these varied preferences. Specifically, we employ the residual vector quantization technique, enabling the interest histogram to be learned in an end-to-end manner.

 As illustrated in Figure~\ref{fig:framework}, after processing the user input  $\{\e_{t}^{\u}\}_{t=1}^{T}$ through the Interest Representation Module, we obtain the input sequence of the Interest Compression Module, denoted by $\{\b_t\}_{t=1}^T$. The ICM utilizes a multi-layer residual quantization structure, where the codebook of each layer quantizes the residuals from the preceding layer.
For the $m$-th layer of the ICM, a codebook $\mathbf{C}^{(m)} \in \R^{K_{m} \times D_{m}}$ is maintained, containing $K_{m}$ code vectors, each with a dimensionality of $D_{m}$. For each input representation $\b_t^{(m)}$ of the $m$-th codebook, we determine the  corresponding code index $k_t^{(m)} \in [K_m]$ and the code vector $\c_t^{(m)} \in \R^{D_{m}}$ through a nearest neighbor search. 
Then, the input to the next codebook is the residual between the input $\b_t^{(m)}$ and code vector $\c_t^{(m)}$, denoted by $\b_t^{(m+1)} = \b_t^{(m)} - \c_t^{(m)}$. 

\vspace{1mm}
\noindent\textbf{Training Stage.} 
Based on the code vector $\c_t^{(m)}$, we can construct a straight-through estimator: 
$\hat{\b}_t^{(m)} = \b_t^{(m)} + \operatorname{sg}\left(\c_t^{(m)} - \b_t^{(m)}\right),$
where $\operatorname{sg}$ denotes the stop-gradient operation. The final user hidden vector $\z_t$ is computed as the sum of all estimators $\sum_{m=1}^{M} \hat{\b}_t^{(m)}$, which is then passed through the user GLU, where $M$ denotes the number of codebooks.
To encourage closer alignment between the representations and their corresponding codes, we minimize the commitment loss, denoted by 
$\mathcal{L}_{\text {commit }}=\sum_{m=1}^M \left\|\mathbf{b}_t^{(m)}-\operatorname{sg}\left(\mathbf{c}_t^{(m)}\right)\right\|_2^2.$ Additionally, the codebook is updated using an exponential moving average strategy.

\vspace{1mm}
\noindent\textbf{Inference Stage.} 
During inference, given a user input sequence $\{\e_{t}^{\u}\}_{t=1}^{T}$, a code index sequence $\{k_t^{(m)}\}_{t=1}^T$ is computed for the $m$-th codebook, where $k_t^{(m)}$ represents the nearest code index for the position $t$. Based on this code index sequence $\{k_t^{(m)}\}_{t=1}^T$, a histogram $\x^{(m)} \in \R^{K_l}$ is constructed, where the $k$-th element corresponds to the frequency of the code index $k$. By concatenating histograms from all codebooks, the final compressed vector is generated, which serves as input for downstream ranking models.
%!TEX root = ../main.tex

% \usepackage{multirow}
% \usepackage{graphicx}
% \usepackage{multirow}
% \usepackage{graphicx}
% \usepackage{multirow}
% \usepackage{graphicx}
\begin{table}[!t]
\centering
% \vspace{-3mm}
\caption{Dataset statistics. Avg. Len. and P90 Len. indicate the average and the 90th percentile length of historical behaviors.}
\label{tab:stats}
\vspace{-3mm}
\resizebox{\columnwidth}{!}{%
\begin{tabular}{ccccccc}
\toprule
\multirow{2}{*}{\textbf{Dataset}} &
  \multirow{2}{*}{\textbf{Partition}} &
  \multirow{2}{*}{\textbf{\#Users}} &
  \multirow{2}{*}{\textbf{\#Items}} &
  \multirow{2}{*}{\textbf{\#Samples}} &
  \multirow{2}{*}{\textbf{Avg. Len.}} &
  \multirow{2}{*}{\textbf{P90 Len.}} \\
                                     &                &       &         &          &     &     \\ \midrule 
\multirow{2}{*}{\textbf{MicroVideo}} & \textbf{Train} & 10887 & 1006093 & 6346207  & 170 & 322 \\
                                     & \textbf{Test}  & 8160  & 451101  & 1427698  & 265 & 457 \\ \midrule
\multirow{2}{*}{\textbf{Tmall}}      & \textbf{Train} & 13610 & 1447822 & 7242504  & 86  & 147 \\
                                     & \textbf{Test}  & 13213 & 952601  & 1366298  & 124 & 198 \\ \midrule
\multirow{2}{*}{\textbf{EBNeRD}}     & \textbf{Train} & 66420 & 7873    & 45823885 & 83  & 132 \\
                                     & \textbf{Test}  & 62736 & 6165    & 9377940  & 121 & 179 \\ \bottomrule
\end{tabular}%
}
\vspace{-3mm}
\end{table}
\subsection{Discussion of Efficiency}
\label{sec:efficiency}

The proposed algorithm exhibits notable training and inference efficiency across three key aspects.  (1)~\textbf{Framework}: The proposed compression-based framework precomputes user representations offline, thereby eliminating the need for repeated online retrieval in the search-based framework~\citep{CIKM'20/SIM, Sigir'20/UBR4CTR} and reduces online computational latency; (2)~\textbf{Compression task}: By representing each behavior as a single token, the compression model reduces computational complexity compared to text-based methods~\citep{NAACL'24/Align, DBLP:conf/www/ZhengCQZ024} that require multiple tokens. Moreover, user representations are obtained through a single forward pass, avoiding the sequential decoding in text-based models. Besides, the autoregressive framework with the holistic loss enables to compute $\O(T^2)$ loss terms from a sequence of length $T$ in a single forward pass, which is more efficient than methods that calculate losses only between certain partitions~\citep{KDD'22/PinnerFormer, arxiv'24/LEARN}; (3)~\textbf{Model architecture}: The compression model shares the backbone architecture of LLMs, facilitating seamless integration with training and inference optimization techniques, such as FlashAttention~\citep{DBLP:conf/nips/DaoFERR22} and DeepSpeed~\citep{DBLP:conf/kdd/RasleyRRH20}.

% The proposed algorithm demonstrates significant advantages in training and inference efficiency. 
% By representing each behavior as a single token, it reduces computational complexity and memory usage compared to text-based methods~\citep{NAACL'24/Align, DBLP:conf/www/ZhengCQZ024} that require multiple tokens. User representations are obtained through a single forward pass, avoiding the sequential decoding in text-based models. Besides, the autoregressive framework with the holistic loss enables to compute $\O(T^2)$ loss terms from a sequence of length $T$ in one forward pass, which is more efficient than methods that calculate losses only between certain partitions~\citep{KDD'22/PinnerFormer, arxiv'24/LEARN}.
% Additionally, the compression model shares the architecture of LLMs, facilitating seamless integration with training and inference optimizations such as FlashAttention~\citep{DBLP:conf/nips/DaoFERR22} and DeepSpeed~\citep{DBLP:conf/kdd/RasleyRRH20}. 
% Most importantly, given the stability of long-term user interests,  the compression model can be updated less frequently (e.g., daily or weekly), which further reduces resource demands. Precomputed user representations can be stored and directly accessed, resulting in no significant impact on online inference efficiency.
%!TEX root = ../main.tex

\section{Experiments}
\label{sec:experiment}
This section presents experimental results. Section~\ref{sec:setup} details experimental setup. Section~\ref{sec:performance} demonstrates the overall performance on three datasets.  Section~\ref{sec:ablation} conducts ablation studies.

\subsection{Experimental Setup} 
\label{sec:setup}

% \vspace{2mm}
\noindent\textbf{Datasets.}
To evaluate the proposed approach, we use three public datasets covering multiple domains and modalities, including both CTR and CVR estimation. The dataset details are as follows:

\begin{itemize}[leftmargin=3mm]
    \item \textbf{MicroVideo}~\citep{DBLP:conf/mm/ChenLZZXL18} is a short video dataset for CTR  prediction. The 64-dimensional representation of cover images is extracted via the Inception-v3 model~\citep{DBLP:conf/cvpr/SzegedyVISW16}.
    \item \textbf{Tmall}~\citep{DBLP:journals/corr/ZhangPSW14} is an e-commerce dataset. We use it for CVR estimation, predicting if a user adds the product to the cart after a click. We download product images from URLs and use the CLIP model~\citep{DBLP:conf/icml/RadfordKHRGASAM21} to extract 256-dimensional representations. 
    \item \textbf{EBNeRD}~\citep{DBLP:conf/recsys/KruseLKPP0UAF24a} is a news recommendation dataset from Ekstra Bladet impression logs, used for CTR estimation. We use the 768-dimensional article representations drawn by the Bert model~\citep{DBLP:conf/naacl/DevlinCLT19}.
\end{itemize}

For each dataset, we select users with long behaviors and split the logs by two time thresholds $t_1$ and $t_2$, which are chosen based on the data distribution. Logs before $t_1$ form the \emph{representation set} for training compression models. Logs in $[t_1, t_2)$ form the \emph{training set} for training utilization models, and later logs form the \emph{test set} for evaluation. Notably, since the compression model only uses logs before $t_1$, it leverages fewer behaviors than the utilization model. This can be improved by splitting more time slices and updating the compression model more frequently, so the actual performance of CHIME is expected to be better than reported.

For each sample in the training and test sets, the input features consist of categorical attributes such as item ID, user ID, and category ID, as well as the behavior sequence comprising all positive historical actions. Specifically, for a sample with timestamp $t$, the historical behavior sequence includes all actions from the same user with timestamps in the range $(-\infty, t)$. The detailed statistics for each dataset are shown in Table~\ref{tab:stats}.

%!TEX root = ../main.tex

\begin{table*}[]
\centering
% \vspace{5mm}
\caption{The overall results of CHIME on three datasets ($\%$). We report the average results over 5 rounds. Boldface refers to the highest
score. "$^{\ast}$" indicates the statistically significant improvements over the baseline by the paired t-test at a $5\%$ significance level. }
\label{tab:overall}
\begin{tabular}{cccccccccc}
\toprule
 & \multicolumn{3}{c}{\textbf{MicroVideo}}         & \multicolumn{3}{c}{\textbf{Tmall}}           & \multicolumn{3}{c}{\textbf{EBNeRD}}           \\ \cmidrule{2-10}  
 & \textbf{AUC\;($\uparrow$)} & \textbf{GAUC\;($\uparrow$)} & \textbf{LogLoss\;($\downarrow$)} & \textbf{AUC\;($\uparrow$)} & \textbf{GAUC\;($\uparrow$)} & \textbf{LogLoss\;($\downarrow$)} & \textbf{AUC\;($\uparrow$)} & \textbf{GAUC\;($\uparrow$)} & \textbf{LogLoss\;($\downarrow$)} \\ 
 \midrule
\textbf{DIN} &
  74.19 &
  69.28 &
  40.10 &
  71.26 &
  59.64 &
  32.57 &
  80.64 &
  80.73 &
  25.35 \\
\textbf{DIN w/ LEARN} &
   74.22 &
  69.34 &
  40.08 &
   71.32 &
   \textbf{59.68} &
   32.54 &
   81.16 &
   81.35 &
   25.83 \\
\textbf{DIN w/ CHIME} &
  $\textbf{74.36}^{\ast}$ &
  $\textbf{69.42}^{\ast}$ &
  $\textbf{40.02}^{\ast}$ &
  \textbf{71.37} &
  59.66 &
  \textbf{32.53} &
  $\textbf{81.42}^{\ast}$ &
  $\textbf{81.43}^{\ast}$ &
  \textbf{24.31} \\ \midrule
\textbf{SimSoft} &
  74.53 &
  69.86 &
  39.87 &
  71.30 &
  59.57 &
  32.56 &
  80.88 &
  80.89 &
  24.78 \\
\textbf{SimSoft w/ LEARN} &
   74.58 &
   \textbf{69.94} &
   39.87 &
   71.31 &
   59.67 &
   32.55 &
   81.42 &
   81.29 &
   \textbf{23.70} \\
\textbf{SimSoft w/ CHIME} &
  $\textbf{74.67}^{\ast}$ &
  69.93 &
  $\textbf{39.81}^{\ast}$ &
  $\textbf{71.42}^{\ast}$ &
  \textbf{59.70} &
  $\textbf{32.51}^{\ast}$ &
  $\textbf{81.70}^{\ast}$ &
  $\textbf{81.61}^{\ast}$ &
  23.85 \\ \midrule
\textbf{ETA} &
  74.40 &
  69.69 &
  39.95 &
  71.33 &
  \textbf{59.69} &
  32.55 &
  80.76 &
  80.83 &
  25.90 \\
\textbf{ETA w/ LEARN} &
   74.43 &
    69.75 &
   39.92 &
   71.31 &
   59.66 &
   32.54 &
   81.15 &
   81.18 &
   24.55 \\
\textbf{ETA w/ CHIME} &
  $\textbf{74.56}^{\ast}$ &
  \textbf{69.76} &
  $\textbf{39.88}^{\ast}$ &
  $\textbf{71.40}^{\ast}$ &
  \textbf{59.69} &
  $\textbf{32.52}^{\ast}$ &
  $\textbf{81.44}^{\ast}$ &
  $\textbf{81.47}^{\ast}$ &
  $\textbf{24.18}^{\ast}$ \\ \midrule
\textbf{SDIM} &
  74.40 &
  69.69 &
  \textbf{39.86} &
  71.18 &
  59.38 &
  32.61 &
  80.76 &
  80.71 &
  24.18 \\
\textbf{SDIM w/ LEARN} &
   74.46 &
   69.80 &
   39.87 &
    \textbf{71.20} &
    59.32 &
    \textbf{32.59} &
   81.00 &
   81.00 &
   24.54 \\
\textbf{SDIM w/ CHIME} &
  $\textbf{74.56}^{\ast}$ &
  $\textbf{69.84}^{\ast}$ &
  \textbf{39.86} &
  \textbf{71.20} &
  \textbf{59.41} &
  32.62 &
  $\textbf{81.22}^{\ast}$ &
  $\textbf{81.09}^{\ast}$ &
  \textbf{24.11} \\ \midrule
\textbf{TWIN} &
  74.79 &
  70.26 &
  39.84 &
  71.33 &
  59.70 &
  32.54 &
  80.69 &
  80.79 &
  24.49 \\
\textbf{TWIN w/ LEARN} &
   74.81 &
   70.27 &
   39.81 &
   71.35  &
    59.71 &
    32.53 &
   81.34 &
   81.33 &
   24.36 \\
\textbf{TWIN w/ CHIME} &
  $\textbf{74.91}^{\ast}$ &
  \textbf{70.30} &
  $\textbf{39.78}^{\ast}$ &
  $\textbf{71.43}^{\ast}$ &
  \textbf{59.75} &
  $\textbf{32.51}^{\ast}$ &
  $\textbf{81.42}^{\ast}$ &
  $\textbf{81.39}^{\ast}$ &
  \textbf{23.97} \\ \bottomrule
\end{tabular}
\vspace{-3mm}
\end{table*}

\vspace{1mm}
\noindent\textbf{Baselines.} Since CHIME produces plug-and-play representations, we integrate them directly into various sequence modeling methods for comparison, including:

\begin{itemize}[leftmargin=3mm]
    \item \textbf{DIN}~\citep{DBLP:conf/kdd/ZhouZSFZMYJLG18} is the most widely used behavior modeling approach, which leverages target attention to model the relationship between candidate items and the short-term behavior sequence.
    \item \textbf{SimSoft}~\citep{CIKM'20/SIM} is a search-based long-term modeling algorithm, where GSU and ESU are trained simultaneously, and the top-k behaviors are chosen via maximum inner product search.
    \item \textbf{ETA}~\citep{arxiv'21/ETA} uses locality-sensitive hashing (LSH) and Hamming distance to accelerate the search process.
    \item \textbf{SDIM}~\citep{CIKM'22/SDIM} gathers behavior items associated with the candidate item with the same hash signature to form the user interest. 
    \item \textbf{TWIN}~\citep{KDD'23/TWIN} unifies the parameters of GSU and ESU, periodically synchronizing the parameters of ESU to GSU.
\end{itemize}

We compare the following behavior compression methods:
\begin{itemize}[leftmargin=3mm]
    \item \textbf{LEARN}~\citep{arxiv'24/LEARN} uses the pretrained LLM to extract item content embeddings and project them to the collaborative space via the dense all action loss. In our experiments, we replace the content embeddings with multi-modal embeddings. 
    \item \textbf{CHIME} is the proposed  behavior compression approach. 
\end{itemize}

% \vspace{2mm}
\noindent \textbf{Metrics.} We adopt three widely used metrics: AUC, gAUC, and log loss. The AUC is a standard metric for binary classification tasks by calculating the area under the receiver operating characteristic (ROC) curve, where a higher AUC indicates better model performance. The gAUC extends the traditional AUC by grouping predictions based on users, allowing us to better understand how well the model performs across different user segments. Logloss measures the accuracy of the predicted probabilities.

\vspace{1mm}
\noindent \textbf{Implementations.} We use the LLaMA-Factory library~\citep{DBLP:journals/corr/abs-2403-13372} to implement the compressed model and FuxiCTR libiary~\citep{DBLP:conf/cikm/ZhuLYZH21, DBLP:conf/sigir/ZhuDSMLCXZ22} to implement the utilization model. The details are listed as follows:
\begin{itemize}[leftmargin=3mm]
    \item \textbf{Compression model.} We use Qwen2.5-0.5B~\citep{qwen2.5} as the backbone model, initializing with the pre-trained parameters of its 24 decoder layers. During training, a learning rate of $7\times 10^{-4}$ is applied to randomly initialized parameters, while a learning rate of $7\times 10^{-6}$ is used for parameters initialized from the pre-trained model. The learning rate is adjusted via a cosine annealing schedule combined with a warm-up strategy over 50 steps. The model is trained on eight A800 GPUs, with a batch size of 8 per GPU. Gradient accumulation with a step size of 4 is employed. The Interest Extractor is configured with 3 layers, where the number of codes in each layer is set to $(32, 16, 16)$. We employ the rotation trick~\citep{DBLP:journals/corr/abs-2410-06424} to mitigate the codebook collapse issue. The temperature coefficients for the holistic and immediate losses are set to 0.2 and 0.07, respectively.
    \item \textbf{Utilization model.} For all methods, the maximum length of short-term behaviors is set to 30, and that of long-term behaviors is set to 1000, covering most samples in the dataset. The learning rate is uniformly set to $5\times 10^{-4}$. The batch size is set to 2048. And the early stopping mechanism is adopted. Hash numbers of SDIM is set to 48 and the width parameter is set to 3. The synchronization frequency of TWIN from ESU to GSU is set to 100 training batches. All methods use the corresponding multimodal embeddings, allowing multimodal information to be utilized even in the baselines, which has been proven to significantly improve performance compared to using only ID features~\citep{DBLP:conf/cikm/GeZZCLYHLSLYHZZ18, DBLP:journals/corr/abs-2306-05001, DBLP:conf/cikm/ShengYGWCZCZG0J24}.
\end{itemize}

\subsection{Overall Performance} 
\label{sec:performance}
As shown in Table~\ref{tab:overall}, in comparison to DIN that only uses short-term behaviors, the introduction of CHIME results in significant improvements across all datasets, indicating that the proposed CHIME approach effectively models long-term interests. Despite using older and less behavioral data than search-based methods, CHIME still achieves competitive results on several datasets (e.g., Tmall, EBNeRD), suggesting further gains with more time slices and frequent updates. When CHIME is integrated into search-based long-term modeling methods, it still provides notable gains, demonstrating that CHIME compensates for the behavioral information loss in search-based approaches.
Furthermore, CHIME shows a clear advantage over other behavioral compression methods, such as LEARN, highlighting its superior design in compression and interest encoding.

Overall, CHIME has demonstrated effectiveness across short video, e-commerce, and news recommendation domains, yielding positive results for both CTR and CVR tasks. It can be easily integrated with other sequence modeling methods, further reinforcing the versatility and applicability.

\subsection{Ablation Studies} 
\label{sec:ablation}
In this section, we perform ablation studies on the MicroVideo dataset to verify the effectiveness of the proposed approach. Specifically, we focus on the following three questions:

\noindent$\bullet$~\textsf{Q1}: Does pre-trained LLM initialization improve performance?

\noindent$\bullet$~\textsf{Q2}: Do the model components (IF module, IE module, next action loss) contribute effectively?

\noindent$\bullet$~\textsf{Q3}: Has the interest encoding module effectively captured the interest distribution?
%!TEX root = ../main.tex

% \begin{table}[t!]
% \centering
% \caption{(\%)}
% \label{tab:my-table}
% \begin{tabular}{cccc}
% \toprule
% \textbf{}            & \textbf{AUC ($\uparrow$)} & \textbf{GAUC ($\uparrow$)} & \textbf{LogLoss ($\downarrow$)} \\ \midrule
% \textbf{CHIME}        & 74.36 $\pm$ 0.03 & 69.42 $\pm$ 0.04  & 40.02 $\pm$ 0.01     \\
% \textbf{CHIME w/o AL} & 74.27 $\pm$ 0.03 & 69.39 $\pm$ 0.05  & 40.08 $\pm$ 0.02     \\
% \textbf{CHIME w/o RQ} & 74.32 $\pm$ 0.02 & 69.34 $\pm$ 0.03  & 40.02 $\pm$ 0.02     \\
% \textbf{CHIME w/o MM} & 74.32 $\pm$ 0.02 & 69.33 $\pm$ 0.03  & 40.05 $\pm$ 0.01     \\ \bottomrule
% \end{tabular}
% \end{table}

\begin{table}[t!]
\centering
\vspace{1mm}
\caption{Ablation study. Performance ($\%$) after removing Immediate Loss (IL), Interest Compression Module (ICM), and Interest Adaptation Module (IAM).}
\label{tab:ablation-module}
\begin{tabular}{cccc}
\toprule
\textbf{}            & \textbf{AUC ($\uparrow$)} & \textbf{GAUC ($\uparrow$)} & \textbf{LogLoss ($\downarrow$)} \\ \midrule
\textbf{CHIME}        & \textbf{74.36} & \textbf{69.42}  & \textbf{40.02}      \\
\textbf{CHIME w/o IL} & 74.27  & 69.39   & 40.08      \\
\textbf{CHIME w/o ICM} & 74.32  & 69.34   & \textbf{40.02}     \\
\textbf{CHIME w/o IAM} & 74.32  & 69.33   & 40.05     \\ \bottomrule
\end{tabular}
\vspace{-4mm}
\end{table}

\begin{figure}[!t]  %子图加并列
	\centering
	\vspace{2mm}
    \includegraphics[width=.75\linewidth]{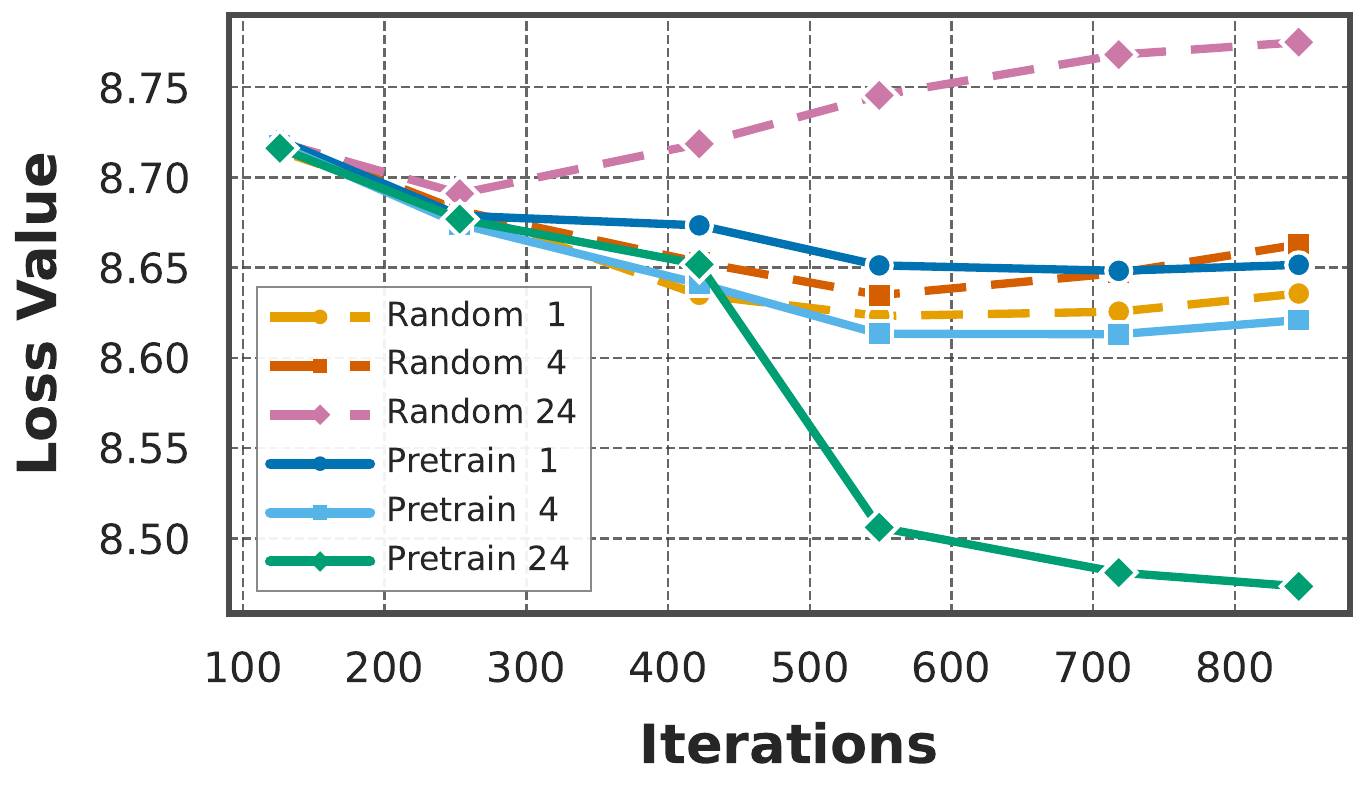}

    \vspace{-2mm}
	\caption{Changes of the evaluation holistic loss values during training. For random initialization, deeper layers lead to more severe overfitting; whereas, for pretrained LLM initialization, deeper layers result in better performance. Overall, pretrained LLM initialization outperforms random initialization.}
	\vspace{-5mm}
	\label{fig-nce}
\end{figure}

\subsubsection{Analysis of Pre-trained LLM Initialization (Q1)}
\label{sec-ablation-llm}
To answer Q1, we compare the performance of the compression model with randomly initialized parameters and parameters initialized using the pre-trained LLM under different numbers of decoder layers. Figure~\ref{fig-nce} illustrates the change in the evaluation loss of the compression model over the training process. From the Figure~\ref{fig-nce}, it is evident that the evaluation loss for the pre-trained LLM initialization is generally higher than that for randomly initialized model, which demonstrates that pre-trained LLM initialization is superior to the randomly initialization. 

When the decoder consists of only one layer, the random initialization slightly outperforms the pre-trained initialization in terms of evaluation loss. However, as the number of layers increases, the evaluation loss for the random initialization progressively worsens, reflecting significant overfitting observed when the decoder has 24 layers. This overfitting is reflected in the increasing evaluation loss as training progresses. In contrast, the pre-trained LLM initialization shows the opposite trend: as the number of layers increases, the evaluation loss steadily decreases, and a significant advantage is observed when the model reaches 24 layers. This demonstrates the effectiveness of using pre-trained LLM parameters in enhancing the compression model's performance, particularly as the model complexity grows.

Further comparison was made to assess the performance of the user representations produced by the compression model when incorporated into the DIN algorithm for the CTR task. As shown in Table~\ref{tab:ablation-llm}, with the increase in the number of decoder layers, the AUC and GAUC for the random initialization progressively decrease, indicating a decline in performance. In contrast, the pre-trained LLM initialization shows the opposite trend, with both AUC and GAUC improving as the decoder layers increase. When the decoder consists of 24 layers and pre-trained initialization is used, the model achieves optimal performance, thereby validating the necessity of initializing the model with pre-trained LLM parameters. This demonstrates the effectiveness of utilizing pre-trained LLM parameters to enhance the performance of the compression model.

\subsubsection{Comparison of Components (Q2)}
To further validate the contributions of each module in the compression model, we evaluated the performance of the CTR task after removing the immediately loss (IL), Interest Compression Module (ICM), and Interest Adaptation Module (IAM). Since removing the Interest Compression Module eliminates the ability to obtain the codebook-based interest distribution, we instead used a 64-dimensional vector derived by applying a GLU to reduce the dimensionality of the LLM Decoder output, followed by average pooling, to serve as the user representation. As shown in Table~\ref{tab:ablation-module}, the removal of the NAP loss from the compression model resulted in a significant drop in AUC for the CTR task. Additionally, removing either the Interest Adaptation Module or the Interest Compression Module lead to decreases in AUC and gAUC for the CTR task, highlighting the critical role these modules play in the overall performance. These findings underscore the importance of the modular design in effectively capturing and utilizing user interest representations.
%!TEX root = ../main.tex

% \usepackage{multirow}
\begin{table}[]
\centering
% \vspace{5mm}
\vspace{-3mm}
\caption{Performance on the CTR task with random initialization and pretrained LLM initialization.
Pretrained LLM initialization with deep layers achieves the best performance.}
\label{tab:ablation-llm}
\begin{tabular}{ccccc}
\toprule
\multicolumn{1}{l}{} &  & \multicolumn{1}{c}{\textbf{AUC ($\uparrow$)}} & \multicolumn{1}{c}{\textbf{GAUC ($\uparrow$)}} & \multicolumn{1}{c}{\textbf{LogLoss ($\downarrow$)}} \\ \midrule
\multirow{2}{*}{\textbf{L=1}}  & \textbf{Random}  & 74.30          & 69.29          & 40.04          \\
                               & \textbf{Pretrain} & 74.28          & 69.28          & 40.08          \\ \midrule
\multirow{2}{*}{\textbf{L=4}}  & \textbf{Random}  & 74.29          & 69.32          & 40.07          \\
                               & \textbf{Pretrain} & 74.30          & 69.32          & 40.07          \\ \midrule
\multirow{2}{*}{\textbf{L=24}} & \textbf{Random}  & 74.26          & 69.25          & 40.07          \\
                               & \textbf{Pretrain} & \textbf{74.36} & \textbf{69.42} & \textbf{40.02} \\ \bottomrule
\end{tabular}
\vspace{-1mm}
\end{table}
%!TEX root = ../main.tex
% \usepackage{graphicx}
\begin{table}[]
\centering
% \vspace{-3mm}
\caption{Hit Ratio and Perplexity of each codebook in the Interest Compression Module.}
\vspace{-3mm}
\label{tab:rqvae}
\resizebox{\columnwidth}{!}{%
\begin{tabular}{cccc}
\toprule
                               & \textbf{Codebook1} & \textbf{Codebook2} & \textbf{Codebook3} \\ \midrule
\textbf{Hit Ratio}             & 91.45\%             & 100\%                  & 100\%                  \\
\textbf{Perplexity / Codesize} & 18.04 / 32         & 14.80 / 16         & 15.40 / 16         \\ \bottomrule
\end{tabular}%
}
% \vspace{-5mm}
\end{table}

\subsubsection{Visualization of Intereset Encoding (Q3)}
\label{sec-visualization}
% question 1

To address Q3, we conduct an analysis of the interest compression module. Table~\ref{tab:rqvae} presents the hit ratio and perplexity for each codebook. The hit ratio is defined as the proportion of codes in the codebook that are activated during the encoding process, reflecting the diversity of code utilization. A higher hit ratio indicates that more codes are actively contributing to the representation, signifying better codebook utilization. The perplexity measures the average uncertainty in code assignments, defined as $2^H$, where $H$ is the entropy of the code distribution. Its range is determined by the size of the codebook, with higher perplexity indicating more evenly distributed code usage and lower perplexity implying dominance by a few codes.
From Table~\ref{tab:rqvae}, it can be observed that most codes are highly active, with the utilization rate of the last two layers' codebooks reaching $100\%$. This demonstrates the effectiveness of the compression model in fully leveraging the available codebook capacity to encode user behaviors.

\begin{figure}[!t]  %子图加并列
    % \vspace{-3mm}
	\centering
	\includegraphics[width=.75\linewidth]{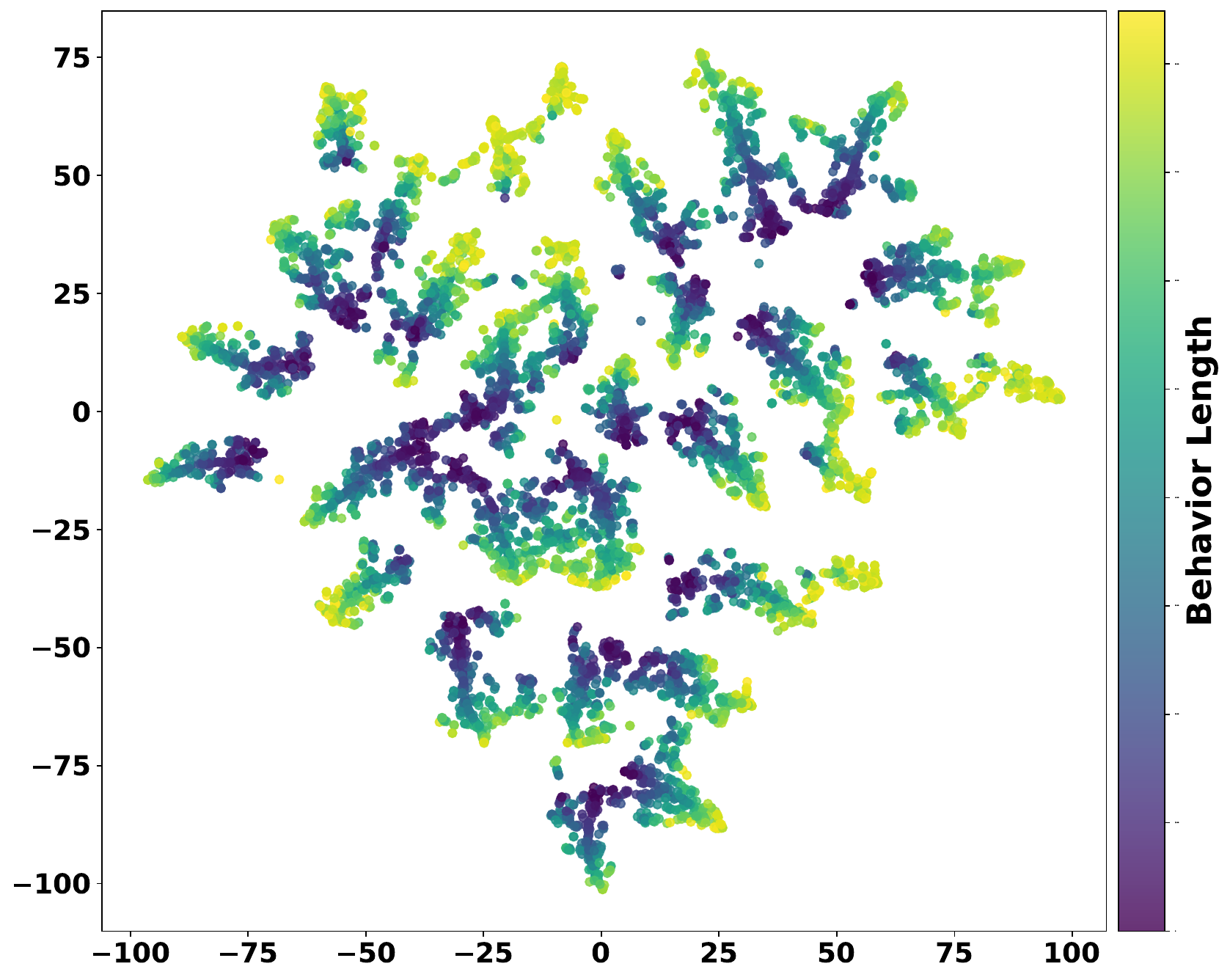}
	\caption{The t-SNE visualization of compressed interest representations reveals distinct clusters, where the colorbar indicates behavior length from purple (shorter) to yellow (longer).}
	\label{fig-tsne}
    \vspace{-3mm}
\end{figure}

Furthermore, Figure~\ref{fig-tsne} provides a t-SNE visualization of the compressed user representations, where the color bar represents the increasing behavior sequence lengths from bottom to top. The visualization reveals a clear clustering structure in the compressed representations, with users of similar behavior lengths grouped closer within the same cluster. This result suggests that the compression model captures user interest distributions with high granularity, effectively representing diverse behavioral patterns and aligning user representations with their behavioral similarities.

%!TEX root = ../main.tex

\section{Conclusion}
\label{sec:conclusion}
In this paper, we proposed a novel approach to compress and utilize long-term user interests, addressing key challenges in representing heterogeneous behavior data.
 To address the challenges of modeling long-term sequences, we introduce an autoregressive compression task and demonstrate the effectiveness of pre-trained LLMs in handling the task, underscoring their suitability for long-term behavior compression. Furthermore, we designed the Interest Adaptation and Interest Compression modules, which efficiently utilize heterogeneous behavioral data to encode and represent user interest distributions. Experimental results on three public datasets demonstrate the effectiveness of our approach in improving CTR and CVR estimation tasks. Furthermore, the insights gained from our approach, including the role of pre-trained parameters as initialization and the representation of user interests via vector quantization, pave the way for future advancements in user modeling and ranking systems.

% \section*{Acknowledgment}
\clearpage
\newpage

\bibliography{ref}
\bibliographystyle{ACM-Reference-Format}

\end{document}